\documentclass[aps,prc,preprint,amsmath,amssymb,nofootinbib,superscriptaddress,longbibliography,floatfix,raggedbottom]{revtex4-2}

\usepackage{graphicx}
\usepackage{xcolor}
\usepackage{tikz}
\makeatletter
\newcommand{\raggedcells}{\let\smc@apbr\@arrayparboxrestore
  \def\@arrayparboxrestore{\smc@apbr\raggedright\let\\\tabularnewline}}
\makeatother
\newcommand{\tablefont}{\def\baselinestretch{1}\footnotesize}
\usetikzlibrary{positioning}

\graphicspath{{fig/}}

\newcommand{\nubar}{\bar\nu}
\newcommand{\GeV}{\,\mathrm{GeV}}
\newcommand{\MeV}{\,\mathrm{MeV}}
\newcommand{\fm}{\,\mathrm{fm}}
\newcommand{\paperI}{Ref.~\cite{Nowak:2026twobody}}
\newcommand{\cm}{\,\mathrm{cm}}
\newcommand{\Lam}{\Lambda}
\newcommand{\Sig}{\Sigma}
\newcommand{\Sst}{\Sigma^{*}(1385)}
\newcommand{\DS}{|\Delta S|}

\providecommand{\slashed}[1]{{\not}#1}

\begin{document}

\title{Phenomenology of strangeness-changing two-body currents\\
in antineutrino scattering}

\author{J.~A.~Nowak}
\affiliation{School of Physics and Astronomy, Lancaster University, Lancaster LA1 4YB, United Kingdom}

\date{\today}

\begin{abstract}
\begin{description}
\item[Background] Antineutrinos produce hyperons without kaons through Cabibbo-suppressed quasi-elastic scattering on a single nucleon. A two-body current acting on a correlated nucleon pair, $\nubar\,[NN]\to\ell^+YN'$, gives the same hyperons with a second nucleon in the final state. A companion paper constructs this current but leaves its absolute size open. StrangeMC, the one event generator that includes the channel, set it to a flat $15\%$ of the quasi-elastic hyperon cross section.
\item[Purpose] To determine the ratio $f=\sigma_{2p2h}/\sigma_{\rm QE}$ of the two-body to
the quasi-elastic hyperon cross section on carbon as a function of antineutrino energy,
and the signatures of two-body events in detectors.
\item[Method] The strangeness-changing two-body current and its non-strange counterpart are computed with the same model on the same nucleon pair. Their
ratio is converted into $f$ with the non-strange two-body (2p2h) and quasi-elastic cross
sections of the Valencia model for carbon. This assumes that the nuclear effects missing from the pair calculation are common to the two currents.
\item[Results] Under this assumption $f$ rises from $0.14$ at $0.6\GeV$ to $0.20$--$0.22$ at $2$--$5\GeV$. Averaged over the Booster antineutrino spectra it is $0.15$--$0.17$, close
to the $15\%$ placeholder. StrangeMC now uses $f(E)$ in place of the placeholder. The
two-body current makes fewer $\Sig^-$ than quasi-elastic production. On proton pairs it
also makes $\Sig^+$ at the weak vertex, which no one-body process can produce. The largest evaluated uncertainty comes from separating the two-body current from one-body production followed by rescattering in the nucleus: across
the separation prescriptions explored, $f$ at $1\GeV$ ranges from $0.12$ to $0.22$.
Interference with one-body production is not calculated and could be as large as the
two-body rate itself.
\item[Conclusions] In a simulated SBND antineutrino-mode exposure on argon that uses the
carbon $f$, two-body events are about $11\%$ of the events that produce a strange hadron. They supply more than a quarter of the kaon-less $\Sig^+$ events used to tag hyperon charge exchange. They reconstruct $12$--$30\%$ low in energy under
the quasi-elastic hypothesis. After final-state interactions half of them have a proton
above $40\MeV$ at the vertex, against a fifth of quasi-elastic events. Hyperon analyses therefore need the channel in their event models. Reducing the uncertainty requires a nuclear calculation of the strange two-body response that includes its interference with one-body production.
\end{description}
\end{abstract}

\maketitle

\section{Introduction}
\label{sec:intro}

Antineutrinos produce hyperons without kaons. On a single nucleon the reactions
$\nubar p\to\ell^+\Lam$, $\nubar p\to\ell^+\Sig^0$ and $\nubar n\to\ell^+\Sig^-$ change
strangeness at the weak vertex. They are Cabibbo suppressed, have no neutrino-induced
counterpart~\cite{singh2006,thorpe2021,benitez2024,fatima2019}, and have been measured on
argon~\cite{microboone2023}. The hyperon is made inside the
nucleus and rescatters before it decays, so these events probe the hyperon--nucleon
interaction in nuclear matter~\cite{companionPRD}.

In non-strange scattering, two-body currents knock out correlated nucleon pairs. This
two-particle--two-hole (2p2h) contribution is a standard part of generator descriptions in
the few-GeV region~\cite{nieves2011,martini2009,amaro2017,schwehr2016,nustec2018,russo2026}.
Antineutrino scattering has a strangeness-changing counterpart,
$\nubar\,[NN]\to\ell^+YN'$, on a correlated pair $[NN]$. Its final state has a hyperon
$Y$, as in quasi-elastic production, and a second nucleon $N'$. Reference~\cite{Nowak:2026twobody} constructs this current from the lowest-order SU(3)
chiral Lagrangian, with contact, meson-exchange, octet-baryon-pole and decuplet-resonance
($\Delta(1232)$, $\Sst$) diagrams. It fixes the relative rates of the seven charge
channels (Sec.~\ref{sec:generator}). One of them makes a $\Sig^+$ on a $pp$ pair, which no
one-body vertex can produce. The calculation is for a
free nucleon pair and has no absolute normalisation. The StrangeMC
generator~\cite{companionPRD} is the only implementation of the channel. It first set the
channel to a flat $15\%$ of the quasi-elastic hyperon cross section, split evenly between
$\Lam$ and $\Sig^0$. Other generators omit the channel, so analyses of hyperon production either
leave it out or use the placeholder. If it is present at the level found here, it adds to
the hyperon rate, changes the hyperon mix and lowers the reconstructed energy of the
events that contain it. It also adds primary $\Sig^+$, made at the weak vertex, to the
samples of $\Sig^+$ without a kaon. These samples are used to study hyperon charge
exchange in the nucleus, such as $\Lam p\to\Sig^+n$.

A generator needs the ratio of the two-body (2p2h) to the quasi-elastic (QE) hyperon cross
section,
\begin{equation}
  f(E) \;\equiv\; \frac{\sigma^{\DS=1}_{2p2h}(E)}{\sigma^{\DS=1}_{\rm QE}(E)}
  \label{eq:fdef}
\end{equation}
where $\DS=1$ marks strangeness-changing reactions. It can be obtained by comparison with ordinary 2p2h, for which nuclear calculations exist. The non-strange
two-body current is computed with the same model on the same pair. The pair calculation
lacks the nuclear density, Fermi motion, Pauli blocking and in-medium modifications. To
the
extent that these act alike on the two currents, they cancel in the ratio $R$ of the two
pair cross sections. The absolute scale then comes from the Valencia calculation of
non-strange 2p2h on carbon~\cite{nieves2011,gran2013}. Carbon is therefore the reference
target, and argon values are an extrapolation.

On carbon $f$ rises from $0.14$ at $0.6\GeV$ to $0.20$--$0.22$ at $2$--$5\GeV$, and its
average over the antineutrino spectra of the Booster Neutrino Beam (BNB) is close to the
$15\%$ placeholder. In simulation, two-body events supply more than a quarter of the
kaon-less $\Sig^+$ events used to study hyperon charge exchange. They differ from
quasi-elastic hyperon events in their reconstructed energy, recoil nucleon, hyperon mix
and muon kinematics. The largest evaluated uncertainty is the treatment of two-body
configurations that amount to one-body production followed by rescattering. The
interference with one-body production is not calculated.

\section{Method}
\label{sec:method}

\subsection{Pair calculation}
\label{sec:mirror}

The non-strange ($\DS=0$) two-body current, called the mirror below, describes ordinary
antineutrino 2p2h, $\nubar\,pn\to\ell^+nn$. Both currents are evaluated on the same
correlated pair: a proton and a neutron with back-to-back momenta of $0.25\GeV$. This is
the lower edge of the momentum range in which short-range-correlated pairs
dominate~\cite{hen2017}. The mirror is the $\DS=1$ current of \paperI\ with nucleon weak vertices in place of the strange ones and
the $\Delta(1232)$ in place of the $\Sst$ (Table~\ref{tab:mirror},
Appendix~\ref{app:current}). The weak $N\to\Delta$ vertices are taken from the
pion-production model of Ref.~\cite{hnv2007}. The two currents share the Lagrangian, the
form factors, the meson cutoffs and the treatment of near-on-shell baryons
(Sec.~\ref{sec:boundary}).

Both currents are restricted to momentum transfers $|\mathbf q|<1.2\GeV$, the range of the
Valencia calculation~\cite{gran2013}. At $0.6$ and $0.8\GeV$ the restriction removes
nothing (Fig.~\ref{fig:qscan}, Appendix~\ref{app:current}). At higher energies it removes
a growing part of the free-pair rate. Without it the $\DS=1$ rate would be $24$--$84\%$
higher at $2$--$5\GeV$ and the mirror rate higher still. $R$ would be $2\%$ lower at
$1\GeV$ and $12\%$ lower at $5\GeV$.

\subsection{Current and cascade}
\label{sec:boundary}

Some two-body diagrams contain an intermediate baryon close to its mass shell.
Such a configuration is a one-body weak reaction followed or preceded by a strong
rescattering on the second nucleon. A generator already simulates it as quasi-elastic
production followed by the intranuclear cascade. It has to be removed from the two-body
current, and how much of it to remove is a choice, called the matching prescription below. The reference choice, a
formation-length cut, removes the configurations in which the intermediate baryon lives
long enough to reach the second nucleon. An intermediate baryon of mass $M$, four-momentum
$p$ and velocity $v$ travels about $2Mv\hbar c/|p^2-M^2|$. Setting this equal to an
internucleon distance of $1.2$--$2.4\fm$ at $v=0.3$--$0.7$ gives a range of cuts on
$|p^2-M^2|$, $\Delta=0.046$--$0.216\GeV^2$, called the formation-length window below
(\paperI). The reference is its centre,
$\Delta=0.10\GeV^2$, which the generator uses. The alternative, a principal-value
treatment of the pole with a smoothing scale $\epsilon=0.02$--$0.06\GeV^2$, suppresses the
near-pole region smoothly instead of removing it. Each prescription is applied identically
to the strange current and to the mirror, so $f$ changes only through the difference in
how the two currents respond. The spread between the prescriptions is the largest evaluated
uncertainty (Sec.~\ref{sec:uncert}). The formulas are in Appendix~\ref{app:current}.

\subsection{Normalisation}
\label{sec:calibration}

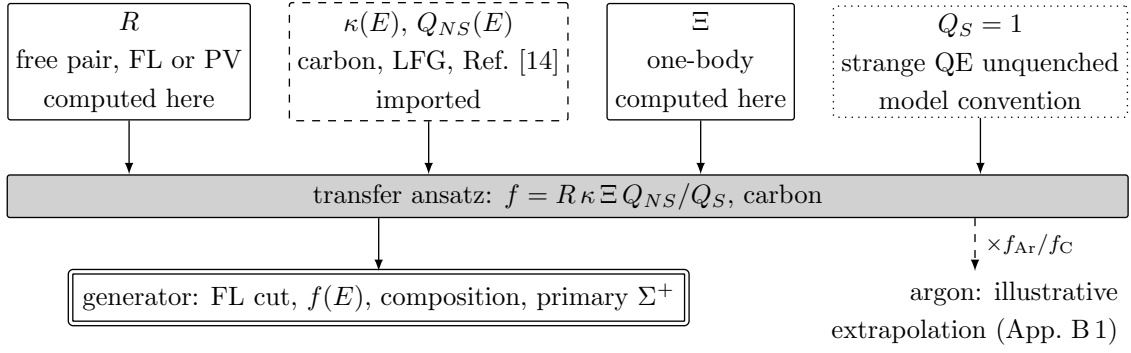
\begin{figure}[t]
\centering
{\def\baselinestretch{1}%
\begin{tikzpicture}[box/.style={draw, rounded corners=1pt, align=center, font=\footnotesize, inner sep=3pt, line width=0.5pt},
  every path/.style={line width=0.5pt}, >=latex]
\node[box] (R) {$R$\\ free pair, FL or PV\\ computed here};
\node[box, right=5mm of R, dashed] (k) {$\kappa(E)$, $Q_{NS}(E)$\\ carbon, LFG, Ref.~\cite{gran2013}\\ imported};
\node[box, right=5mm of k] (X) {$\Xi$\\ one-body\\ computed here};
\node[box, right=5mm of X, dotted] (Q) {$Q_S=1$\\ strange QE unquenched\\ model convention};
\path (R.south west); \pgfgetlastxy{\xa}{\ya}
\path (Q.south east); \pgfgetlastxy{\xb}{\yb}
\path (k.south); \pgfgetlastxy{\xk}{\yk}
\pgfmathsetlengthmacro{\fw}{\xb-\xa}\pgfmathsetlengthmacro{\xm}{(\xa+\xb)/2}
\node[box, fill=black!18, minimum width=\fw, anchor=north] (f) at (\xm,\yk-7mm) {transfer ansatz: $f=R\,\kappa\,\Xi\,Q_{NS}/Q_S$, carbon};
\node[box, below=7mm of f, xshift=-25mm, double, double distance=1.2pt, inner sep=4.5pt] (g) {generator: FL cut, $f(E)$, composition, primary $\Sig^+$};
\draw[->] (R) -- (R |- f.north); \draw[->] (k) -- (k |- f.north); \draw[->] (X) -- (X |- f.north); \draw[->] (Q) -- (Q |- f.north);
\draw[->] (g.north |- f.south) -- (g.north);
\node[below=7mm of f.south east, anchor=north east, font=\footnotesize, align=right] (ar) {argon: illustrative\\ extrapolation (App.~\ref{app:argon})};
\draw[->, dashed] (ar.north |- f.south) -- node[right, font=\scriptsize] {$\times f_{\rm Ar}/f_{\rm C}$} (ar.north);
\end{tikzpicture}}
\caption{\label{fig:chain}The normalisation of Eq.~\eqref{eq:chain}. Solid frames:
computed here; dashed frame: taken from the Valencia calculation (Table~\ref{tab:kappa});
dotted frame: model convention; grey: the product, formed under the transfer ansatz;
double frame: what the generator uses. LFG: local Fermi gas; FL: formation-length cut; PV: principal value
(Sec.~\ref{sec:boundary}). The argon values are an illustrative extrapolation
(Appendix~\ref{app:argon}).}
\end{figure}

The strange two-body cross section is related to the strange quasi-elastic one by four
factors (Fig.~\ref{fig:chain}):
\begin{equation}
  f=\frac{\sigma^{\DS=1}_{2p2h}}{\sigma^{\DS=1}_{\rm QE}}
   \;=\;
  \underbrace{\frac{\sigma^{\DS=1}_{2p2h}}{\sigma^{\DS=0}_{2p2h}}}_{R}
  \;\times\;
  \underbrace{\frac{\sigma^{\DS=0}_{2p2h}}{\sigma^{\DS=0,\,\rm RPA}_{\rm QE}}}_{\kappa}
  \;\times\;
  \underbrace{\frac{\sigma^{\DS=0}_{\rm QE}}{\sigma^{\DS=1}_{\rm QE}}}_{\Xi}
  \;\times\;
  \underbrace{\frac{\sigma^{\DS=0,\,\rm RPA}_{\rm QE}}{\sigma^{\DS=0}_{\rm QE}}}_{Q_{NS}} .
  \label{eq:chain}
\end{equation}
Written with nuclear cross sections, Eq.~\eqref{eq:chain} is an identity. $\kappa$ is the
ratio of the antineutrino 2p2h cross section to the quasi-elastic one with correlations in
the random-phase approximation (RPA). $Q_{NS}$ is the factor by which these correlations
lower, or quench, the quasi-elastic cross section. Both are non-strange and are
taken from the Valencia calculation for carbon~\cite{gran2013} (Table~\ref{tab:kappa},
Appendix~\ref{app:inputs}). $\Xi$, the ratio of the non-strange to the strange
quasi-elastic cross section, is computed on free nucleons with the quasi-elastic current
of Refs.~\cite{singh2006,thorpe2021}, as implemented in \paperI. $R$ is computed on the
free pair. The strange and non-strange sectors use the same code and targets, so
normalisation conventions cancel.

The step from the free pair to the nucleus is the transfer ansatz. Let $\mathcal
N_X=\sigma^X_{2p2h}/\sigma^X_{\rm pair}$ be the factor that turns the free-pair cross
section of sector $X$ into the nuclear one. The ansatz is $\mathcal N_{\DS=1}=\mathcal
N_{\DS=0}$, so that the first factor of Eq.~\eqref{eq:chain} can be computed on the free
pair; this free-pair value is $R$. $\mathcal N_X$ contains everything the pair calculation lacks: the density integration, Fermi motion and Pauli blocking, like-nucleon
pairs, the in-medium resonance self-energy with its absorption part, three-nucleon terms
and the RPA dressing of the two-body response. Section~\ref{sec:uncert} brackets the
resonance absorption and the contribution of like-nucleon pairs.

$Q_{NS}$ converts the quenched quasi-elastic cross section in the denominator of $\kappa$
into the unquenched one in the numerator of $\Xi$. An RPA quenching $Q_S$ of the strange quasi-elastic cross section would
divide the product by $Q_S$; it is set to $1$. This choice rests on the vanishing
$\pi\Lam\Lam$ coupling, which removes the leading pion-exchange part of the quenching for
the $\Lam$. The argument does not hold for the $\Sig$ channels; alternatives are in
Sec.~\ref{sec:uncert}.

\subsection{Benchmark}
\label{sec:valencia}

\begin{figure}[t]
\centering
\includegraphics[width=0.8\textwidth]{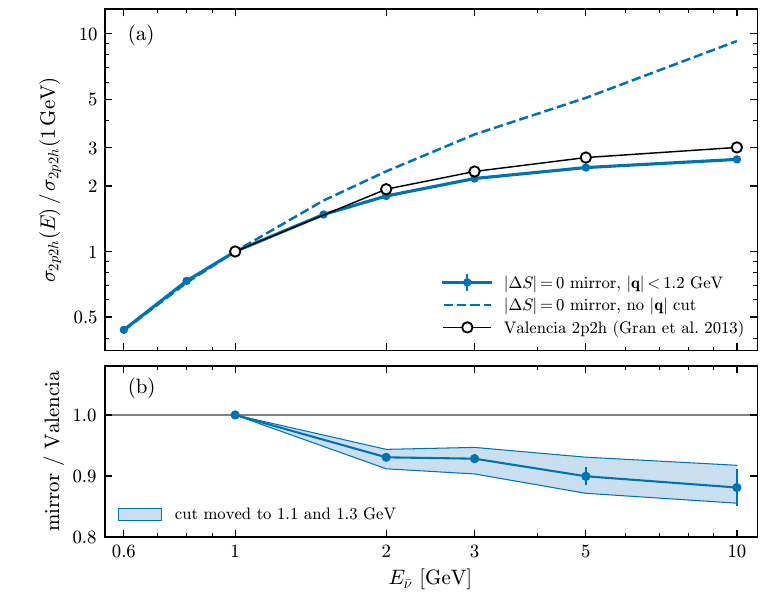}
\caption{\label{fig:valencia}Energy dependence of the non-strange ($\DS=0$) two-body
mirror against the Valencia 2p2h calculation~\cite{gran2013} (antineutrino, carbon),
each normalised at $1\GeV$. (a) The mirror with $|\mathbf q|<1.2\GeV$ (solid; points:
evaluated energies, with Monte Carlo errors), without the restriction (dashed), and the
Valencia values (open circles). (b) Ratio of the restricted mirror to the Valencia
calculation, unity at $1\GeV$ by construction; the band moves the restriction to $1.1$
and $1.3\GeV$.}
\end{figure}

The mirror describes ordinary 2p2h, which the Valencia calculation also gives. Comparing
the two after normalising them at one energy tests the energy dependence of the free-pair
treatment. It does not test the absolute normalisation, which comes from $\kappa$. Gran, Nieves, Sanchez and Vicente
Vacas~\cite{gran2013} extended the model of Ref.~\cite{nieves2011} to $10\GeV$ and
tabulate the antineutrino 2p2h cross section on carbon. Figure~\ref{fig:valencia} compares
its shape with that of the mirror, both normalised at $1\GeV$. Between $1$ and $10\GeV$
the mirror restricted to $|\mathbf q|<1.2\GeV$ rises by a factor $2.65(9)$ and the Valencia calculation by $3.0$,
so the mirror lies $7$--$12\%$ below it at $2$--$10\GeV$. Without the restriction the
mirror rises by $8.6(3)$.

\section{Two-body ratio}
\label{sec:results}

\begin{table}[t]
\caption{\label{tab:chain}The four factors of Eq.~\eqref{eq:chain} and their product at
the formation-length cut $\Delta=0.10\GeV^2$, $|\mathbf q|<1.2\GeV$ and no short-range
correlation. $R$ and $\Xi$ are means of four independent Monte Carlo replicas of $2\times10^4$ phase-space points, with the error on the last digits in parentheses; the error of
$\Xi$ is below $0.2\%$. $\kappa$ and $Q_{NS}$ are from Table~\ref{tab:kappa}; $^\dagger$below $1\GeV$ they are held at their
$1\GeV$ values. $f$ is formed from unrounded inputs, and its error is the replica standard
error of $R\,\Xi$.
A larger set of $27$ runs of $2\times10^4$ phase-space points, which includes the four replicas,
gives
$R$ lower by $0.9$, $2.1$ and $2.0\%$ at $1$, $2$ and $5\GeV$ and within $0.5\%$
elsewhere. The generator uses the four-replica values. The $10\GeV$
row completes the grid for the flux averages.}
\tablefont
\begin{ruledtabular}
\begin{tabular}{rccccc}
$E_{\nubar}$ [GeV] & $R$ & $\Xi$ & $\kappa$ & $Q_{NS}$ & $f$ \\
\hline
0.6$^\dagger$  & 0.0230(1)  & 25.8 & 0.294 & 0.80 & 0.139(1) \\
0.8$^\dagger$  & 0.0308(2)  & 19.8 & 0.294 & 0.80 & 0.143(1) \\
1.0  & 0.0359(3)  & 17.9 & 0.294 & 0.80 & 0.150(1) \\
1.5  & 0.0402(1)  & 16.5 & 0.313 & 0.84 & 0.174(1) \\
2.0  & 0.0426(4)  & 16.2 & 0.331 & 0.88 & 0.202(2) \\
3.0  & 0.0434(15) & 16.1 & 0.328 & 0.92 & 0.211(7) \\
5.0  & 0.0449(13) & 16.2 & 0.324 & 0.95 & 0.223(6) \\
10.0 & 0.0423(14) & 16.3 & 0.320 & 0.96 & 0.213(7) \\
\end{tabular}
\end{ruledtabular}
\end{table}

Table~\ref{tab:chain} gives the four factors and their product under the transfer ansatz,
in the reference configuration: the formation-length cut $\Delta=0.10\GeV^2$, $|\mathbf
q|<1.2\GeV$, $Q_S=1$ and no short-range correlation. $R$ is $0.023$ at
$0.6\GeV$, less than half of the naive Cabibbo expectation
$\tan^2\theta_C=0.053$~\cite{pdg2024}, because the hyperon mass costs phase space near
threshold. It rises to $0.043$--$0.045$ at $2$--$5\GeV$. Over the same range $\Xi$ falls
from $26$ to $16$, and $f$ rises from $0.14$ to $0.22$. Among kaon-less hyperon events from the two
processes, the two-body share $f/(1+f)$ is $12\%$ at $f=0.14$ and $18\%$ at $f=0.22$.
StrangeMC now uses the $f(E)$ of Table~\ref{tab:chain} in place of the flat $15\%$
(Sec.~\ref{sec:generator}).

The results are not equally reliable at all energies. The main range of the calculation is $0.6$--$2\GeV$.
Below $1\GeV$, however, the Valencia table has no entry, and $\kappa$ and $Q_{NS}$ are
held at their $1\GeV$ values. Their product is the 2p2h cross section over the
quasi-elastic one in the local Fermi gas without RPA, $\sigma_{2p2h}/\sigma^{\rm LFG}_{\rm
QE}$ of Table~\ref{tab:kappa}. It falls from $0.293$ at $2\GeV$ to $0.234$ at $1\GeV$. If
it keeps falling below $1\GeV$, the held values overestimate $f$ at $0.6$ and $0.8\GeV$. At $2$--$3\GeV$ the result depends increasingly
on the momentum-transfer restriction (Sec.~\ref{sec:mirror}). At $3$--$5\GeV$, $f$ and the
hyperon composition (Sec.~\ref{sec:generator}) show the behaviour of the free-pair treatment and are not predictions. The
$10\GeV$ point only completes the grid for flux averages.

\begin{figure*}[t]
\centering
\includegraphics[width=0.98\textwidth]{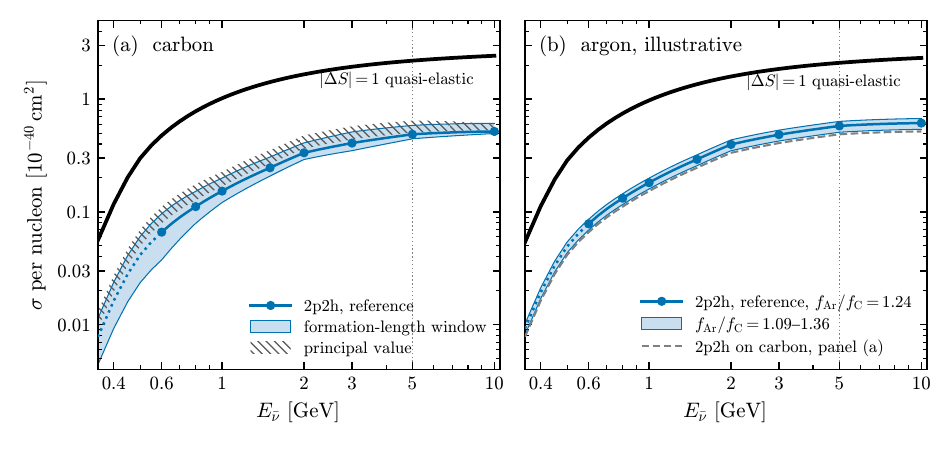}
\caption{\label{fig:sigma}Cross section per nucleon of quasi-elastic hyperon production
(heavy line) and of the strange two-body current, $\sigma_{2p2h}=f\sigma_{\rm QE}$ (thin
line), against antineutrino energy. (a) Carbon: reference (points: evaluated energies),
with the formation-length window (band) and the principal value (hatched) of
Sec.~\ref{sec:uncert}; below $0.6\GeV$ (dotted) $f$ is held at its $0.6\GeV$ value.
(b) Argon, an illustrative extrapolation (Appendix~\ref{app:argon}). The quasi-elastic
cross section is scaled by $0.957$. The two-body cross section uses $f_{\rm Ar}/f_{\rm
C}=1.24$ (line) and $1.09$--$1.36$ (band), which is $1.19$ ($1.04$--$1.30$) times the
carbon two-body curve (dashed grey).}
\end{figure*}

Multiplying $f$ by the strange quasi-elastic cross section gives the two-body cross
section per nucleon (Fig.~\ref{fig:sigma}). On carbon at $1\GeV$ it is
$0.153(1)\times10^{-40}\cm^2$, against $1.018\times10^{-40}\cm^2$ for quasi-elastic hyperon production. The
argon extrapolation of Appendix~\ref{app:argon} puts it at $1.04$--$1.30$ times the
carbon value.

\subsection{Booster averages}
\label{sec:bnb}

\begin{figure}[t]
\centering
\includegraphics[width=0.75\textwidth]{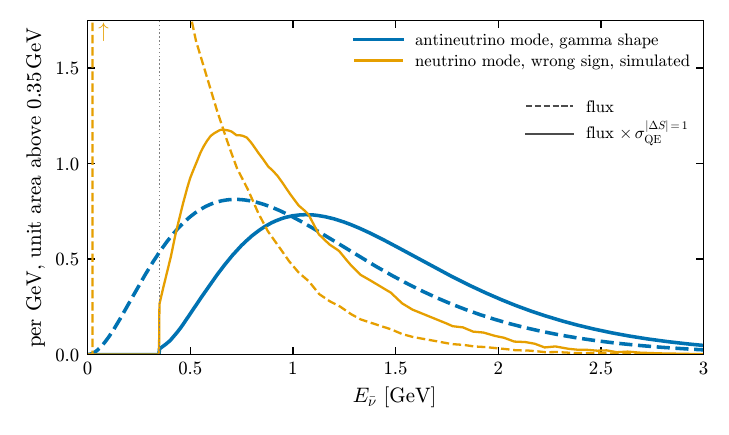}
\caption{\label{fig:flux}The two Booster antineutrino spectra over which $f$ is averaged.
Heavy lines: the antineutrino-mode $\bar\nu_\mu$ spectrum in the gamma-distribution
parametrisation of the generator. Thin lines: the wrong-sign $\bar\nu_\mu$ component of
the neutrino-mode flux at SBND, from the Booster beam simulation of
Ref.~\cite{miniboone2009flux} evaluated at the SBND location. Dashed: flux, normalised to unit area above $0.35\GeV$; the wrong-sign flux
peaks off scale at $0.08\GeV$ (arrow). Solid: rate weight, flux times the $\DS=1$ quasi-elastic
cross section, normalised to unit area over $0.35$--$10\GeV$, the range of the average. The
dotted line marks $0.35\GeV$.}
\end{figure}

A generator that applies one number to a Booster beam should use the average
\[
  \langle f\rangle=\frac{\int_{0.35\GeV}^{10\GeV}dE\,\Phi(E)\,\sigma^{\DS=1}_{\rm QE}(E)\,f(E)}
                        {\int_{0.35\GeV}^{10\GeV}dE\,\Phi(E)\,\sigma^{\DS=1}_{\rm QE}(E)},
\]
the ratio of the flux-integrated two-body and quasi-elastic hyperon rates, with $\Phi$ the
antineutrino flux.
It is $0.169$ for the antineutrino-mode (reverse horn current, RHC) spectrum. For the
wrong-sign $\bar\nu_\mu$ component of the neutrino-mode (forward horn current, FHC) flux
at SBND it is $0.153$ (Fig.~\ref{fig:flux}, Table~\ref{tab:bnb}). Both are close to the
$15\%$ placeholder. The RHC spectrum is approximated by the gamma-distribution
parametrisation that the generator uses. The FHC flux is the Booster beam
simulation of Ref.~\cite{miniboone2009flux} evaluated at the SBND location, $110\,$m from
the target. At the peak of the RHC flux, $0.72\GeV$, $f$ is $0.141$. This is $8$--$17\%$
below both averages, which include the rise of $f$ above $1\GeV$. Both averages hold $f$ at its
$0.6\GeV$ value below $0.6\GeV$.
Energies below $1\GeV$, where $\kappa$ and $Q_{NS}$ are held, carry $28\%$ of the RHC and
$62\%$ of the FHC rate weight (flux times the quasi-elastic hyperon cross section).

\begin{table}[t]
\caption{\label{tab:bnb}What ``the BNB value'' of $f$ can mean, for the reference $f$. The
rate-weight peaks are maxima per unit energy.}
\tablefont
\begin{ruledtabular}
\raggedcells\begin{tabular}{p{3.6cm}cp{6.0cm}}
quantity & value & provenance \\
\hline
$f$ at the grid point $0.6\GeV$ & 0.139 & $R$, $\Xi$ calculated; $\kappa$, $Q_{NS}$ held
at their $1\GeV$ values \\
$f$ at the grid point $1\GeV$ & 0.150 & calculated \\
$f$ at the flux peak, $0.72\GeV$ & 0.141 & interpolated between the $0.6$ and $0.8\GeV$ grid points, both
with $\kappa$ and $Q_{NS}$ held; peak of the RHC spectrum \\
$\langle f\rangle$, RHC spectrum & 0.169 & gamma-distribution parametrisation; rate weight peaks at $1.06\GeV$ \\
$\langle f\rangle$, FHC wrong-sign $\bar\nu_\mu$ & 0.153 & beam simulation at SBND; softer spectrum; rate weight peaks at $0.66\GeV$ \\
\end{tabular}
\end{ruledtabular}
\end{table}

\section{Uncertainties}
\label{sec:uncert}

\begin{figure}[t]
\centering
\includegraphics[width=0.8\columnwidth]{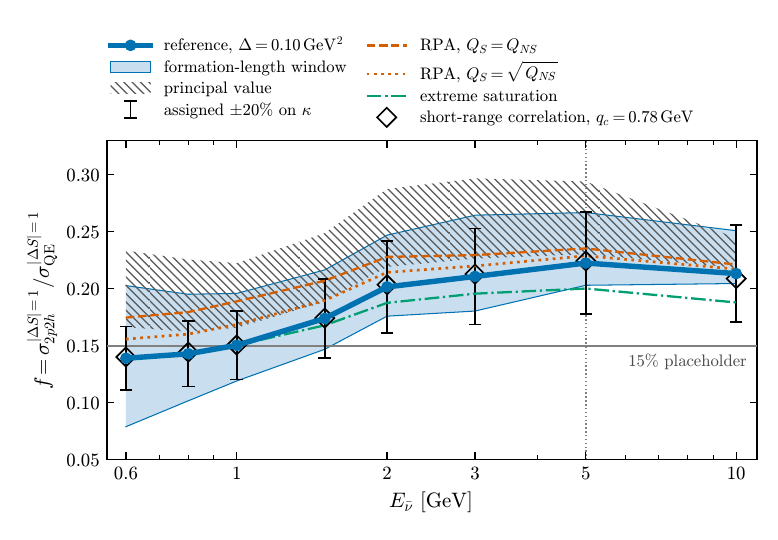}
\caption{\label{fig:hierarchy}$f$ on carbon and its main variations against
antineutrino energy. Solid line and filled circles: reference (formation-length cut
$\Delta=0.10\GeV^2$, $Q_S=1$, no short-range correlation). Band: formation-length window
$\Delta=0.046$--$0.216\GeV^2$; hatched band: principal value,
$\epsilon=0.02$--$0.06\GeV^2$; bars: the $\pm20\%$ assigned to $\kappa$; dashed and
dotted curves: the RPA options $Q_S=Q_{NS}$ and $\sqrt{Q_{NS}}$; open diamonds:
correlation on every meson, $q_c=0.78\GeV$; dash-dotted curve, from $1\GeV$: extreme
saturation; grey line: the $15\%$ placeholder. None of the bands is a probability
interval, and the one-body interference is not shown. Beyond the dotted line at $5\GeV$
the $10\GeV$ point only completes the grid for flux averages.}
\end{figure}

Table~\ref{tab:budget} lists the sensitivity of $f$ to each input, and
Fig.~\ref{fig:hierarchy} shows the main variations against energy. The largest effect, the
current--cascade separation of Sec.~\ref{sec:boundary}, moves the Booster average by a factor $0.80$--$1.47$.
Next come the non-strange input $\kappa$ ($\pm20\%$), the RPA convention (up to $+26\%$) and the
in-medium widths of the decuplet resonances ($-35\%$ to $+15\%$). The other inputs move
$f$ by $10\%$ or less. These are alternative definitions or
assumptions rather than statistical uncertainties, so they are not combined into a
total. The interference with one-body production is not evaluated and could exceed all
of them.

A generator or an analysis that needs one range for $f$ can take the union of the two
prescriptions, $0.80$--$1.47$ times the reference ($\langle f\rangle=0.136$--$0.249$ over
the RHC spectrum). The $\pm20\%$ on $\kappa$ is then a separate normalisation uncertainty.
The interference changes the summed quasi-elastic and two-body hyperon rate. A free
normalisation of that rate absorbs this change, but not the change of the kinematic
distributions.

\paragraph{Current--cascade separation.} Across the formation-length window $f$ at
$1\GeV$ moves between $0.12$ and $0.20$. The principal value gives $1.03$--$1.48$ times
the reference at $1$--$5\GeV$ (Table~\ref{tab:boundary}, Appendix~\ref{app:inputs}).
Near-pole configurations carry a large share of both currents, so each current depends
strongly on the prescription. The ratio $R$ cancels much of this dependence. Over the window the strange pair cross section at $1\GeV$ changes by a
factor $3.4$ (\paperI) and the mirror by $2.0$, while $R$ moves only from $0.79$ to $1.30$
times its reference value. Over the RHC spectrum $\langle f\rangle$ spans $0.136$--$0.217$
across the window and $0.186$--$0.249$ under the principal value. The $15\%$ placeholder
thus lies inside the range spanned by the window and below the principal-value range. The window is not a lower bound: other ways of separating current and cascade could
give smaller values. In every prescription $f$ rises by a factor $1.33$--$1.52$ between
$1$ and $3\GeV$.

\paragraph{Interference with one-body production.} One-body production followed by a
hard rescattering that ejects a second nucleon reaches the same final state as the
two-body current, and the two amplitudes can interfere. The interference is customarily
neglected in the non-strange case, where the two responses peak in different regions. Here
the strange two-body response starts at the hyperon threshold. The octet-baryon-pole
diagrams, which are closest to production followed by rescattering, carry $40$--$54\%$ of
the $\Lam$-channel rate at $1$--$3\GeV$ (\paperI). Let $\rho$, between $-1$ and $1$, be an
effective coherence: a single number for the unknown overlap of the two amplitudes. The
interference term is then
\begin{equation}
  \sigma_{\rm int}=2\rho\sqrt{\sigma_{\rm QE}\,\sigma_{2p2h}},\qquad
  \frac{\sigma_{\rm int}}{\sigma_{\rm QE}}=2\rho\sqrt{f}.
  \label{eq:rho}
\end{equation}
At $1\GeV$, $|\rho|=0.1$ gives $\pm0.08$ of the quasi-elastic rate,
comparable to the spread of $f$ between the two matching prescriptions. At $|\rho|\simeq0.2$ the
interference is as large as the two-body rate. No calculation or measurement
constrains $\rho$. The interference also changes the kinematic distributions. These
changes are not calculated, so a single global $\rho$ is not a parameter a generator can
vary.

\paragraph{Non-strange input and resonance absorption.} Reference~\cite{gran2013}
quotes no uncertainty on $\kappa$. The assigned $\pm20\%$ covers the spread between 2p2h
models~\cite{martini2009,amaro2017,lovato2020} and moves $\langle f\rangle$ over
$0.135$--$0.203$. The transfer ansatz is most exposed through resonance absorption. In
the nucleus a $\Delta(1232)$ can be absorbed on a second nucleon, $\Delta N\to NN$, which
adds 2p2h strength. $\kappa$ contains this absorption, but its strange analogue, $\Sst
N\to YN$, is missing from $R$. That process has not been measured: hypernuclear data constrain $\Lam N\to NN$,
not the absorption of a resonance. The benchmark gives a rough scale: the mirror falls
$7$--$12\%$ short of the Valencia energy dependence at $2$--$10\GeV$
(Sec.~\ref{sec:valencia}). In an extreme scenario, labelled extreme saturation in the
tables, the whole shortfall is non-strange absorption with no strange counterpart. $f$ is
then lower by the same fractions, $7\%$ at $3\GeV$ ($\langle f\rangle=0.164$). If absorption acts alike in the
two sectors, $f$ is unchanged. This is a scenario and not a bound: the shortfall may
have other causes, and normalising at $1\GeV$ hides any difference already present
there.

\paragraph{RPA convention.} Quenching the strange quasi-elastic cross section like the
non-strange one, $Q_S=Q_{NS}$, raises $f$ by $26\%$ at $1\GeV$ and $9\%$ at $3\GeV$. It
raises $\langle f\rangle$ to $0.202$. The intermediate choice $Q_S=\sqrt{Q_{NS}}$ raises $f$ by
$12\%$ and $4\%$ (Table~\ref{tab:rpascen}, Appendix~\ref{app:inputs}).

\paragraph{Decuplet resonances.} The $\Sst$ and $\Delta(1232)$ poles carry a large share
of both currents, and a larger share of the mirror (Appendix~\ref{app:inputs}). In matter both
resonances broaden. Taking the $\Delta(1232)$ width from $117$ to
$200\MeV$~\cite{osetsalcedo1987}, and the $\Sst$ width in proportion, raises $R$ by
$7$--$15\%$. Broadening the $\Sst$ alone to $100\MeV$, within the in-medium broadening of
Ref.~\cite{kaskulov2006}, lowers $R$ by $14$--$35\%$ at $0.8$--$5\GeV$. The weak
$N\to\Delta$ transition form factors come from Ref.~\cite{hnv2007} and are carried to the $\Sst$ by SU(3). A common
change of these form factors would move $R$; it has not been evaluated.

\paragraph{Short-range correlations.} A short-range correlation is a property of the pair
and acts on every meson it exchanges. With the Oset--Toki--Weise correlation
function~\cite{oset1982}, of range parameter $q_c=0.78\GeV$ (Appendix~\ref{app:inputs}),
on every exchanged meson in both currents, each current falls by
about a factor of three. $R$ changes by at most $1\%$ at $0.6$--$5\GeV$, and $\langle
f\rangle$ moves to $0.170$ (RHC).

\paragraph{Smaller effects.} The axial mass, which also changes $\Xi$, moves $f$ by up to
$10\%$ (Table~\ref{tab:budget}). Varying the momentum-transfer restriction between $1.1$
and $1.3\GeV$, the meson cutoff $\Lambda_\pi$ (Appendix~\ref{app:current}), the SU(3)
axial couplings $D$ and $F$, or the induced
pseudoscalar coupling moves $f$ by about $1\%$ or less. $R$ is computed for one pair configuration; its
dependence on the nucleon momenta has not been evaluated.
$f$ is computed from $pn$ pairs. The $^{40}$Ar pair census of \paperI\ gives the relative
numbers of $pp$, $pn$ and $nn$ pairs. In it the like-pair channels, the primary $\Sig^+$
included, add $5\%$ to the strange rate of the $pn$ pairs. The mirror has one like-pair
channel, $\nubar\,pp\to\ell^+pn$, which can only lower $R$. The
census would therefore raise $f$ by at most $5\%$; it is not included.
Appendix~\ref{app:current} discusses the omitted terms of the current (the axial contact
operator and three-nucleon terms) and a form-factor dressing that conserves the vector
current.

\paragraph{Argon.} All values of $f$ are for carbon. Appendix~\ref{app:argon} combines two
published routes to the nuclear dependence of non-strange 2p2h with the one-body cross
sections. They place $f$ on argon $9$--$36\%$ above carbon; for the model that $\kappa$
comes from, $f_{\rm Ar}/f_{\rm C}=1.24$. The published scaling is for neutrinos and the
full response, so this is an illustrative extrapolation and is not included in the carbon
values.

\begin{table*}[tp]
\caption{\label{tab:budget}Sensitivity of $f$ to its inputs. The $1$ and $3\GeV$ columns
give the change relative to the reference: a factor for the current--cascade separation
and a percentage otherwise. Slash-separated entries follow the order of the first column.
The last column is the average over the RHC Booster spectrum
(reference $0.169$; the FHC wrong-sign average is $0.01$--$0.03$ lower in every row).
Rows are re-runs of the calculation with one to four seeds, except $\kappa$, extreme
saturation, pair census and target, which rescale the reference. The one-body
interference [Eq.~\eqref{eq:rho}] is not evaluated.}
\tablefont
\begin{ruledtabular}
\raggedcells\begin{tabular}{p{7.6cm}ccc}
source & $E=1\GeV$ & $3\GeV$ & $\langle f\rangle_{\rm RHC}$ \\
\hline
\multicolumn{4}{l}{\emph{current--cascade separation (alternative definitions)}} \\
formation-length window, $\Delta=0.046$--$0.216\GeV^2$ & $\times0.79$--$1.30$ & $\times0.86$--$1.26$ & 0.136--0.217 \\
principal value, $\epsilon=0.02$--$0.06\GeV^2$ & $\times1.10$--$1.48$ & $\times1.07$--$1.41$ & 0.186--0.249 \\
\multicolumn{4}{l}{\emph{non-strange input and transfer ansatz}} \\
$\kappa$ from Ref.~\cite{gran2013}, assigned range & $\pm20\%$ & $\pm20\%$ & 0.135--0.203 \\
extreme saturation, non-strange absorption only & $0$ & $-7\%$ & 0.164 \\
pair census, like pairs and $\Sig^+$ (not applied) & $\le+5\%$ & $\le+5\%$ & --- \\
\multicolumn{4}{l}{\emph{convention and physics scenarios}} \\
RPA on the strange QE, $Q_S=\sqrt{Q_{NS}}$ / $Q_{NS}$ & $+12$ / $+26\%$ & $+4$ / $+9\%$ & 0.185 / 0.202 \\
in-medium widths, proportional / $\Gamma_{\Sst}$ alone & $+10$ / $-20\%$ & $+9$ / $-28\%$ & --- \\
correlation on every meson, $q_c=0.60$--$1.00\GeV$ & $0$ to $+0.4\%$ & $+0.5$ to $+1.0\%$ & 0.170 ($q_c=0.78$) \\
\multicolumn{4}{l}{\emph{diagnostic}} \\
decuplet deleted, both currents & $+32\%$ & $+29\%$ & 0.222 \\
\multicolumn{4}{l}{\emph{target (illustrative extrapolation, not applied)}} \\
C $\to$ Ar & $+9$ to $+36\%$ & $+9$ to $+36\%$ & 0.210 \\
\multicolumn{4}{l}{\emph{parameters and numerics}} \\
$|\mathbf q|_{\max}=1.1$--$1.3\GeV$ & $+1.0$ / $-1.1\%$ & $-0.5$ / $+0.7\%$ & --- \\
axial mass, $M_A=0.9$--$1.2\GeV$ & $+3$ / $-7\%$ & $+5$ / $-10\%$ & --- \\
meson cutoff, $\Lambda_\pi=1.0$--$1.5\GeV$ & $-0.2$ / $+0.1\%$ & $+0.3$ / $-0.2\%$ & --- \\
$D$, $F$, each $\pm0.008$ & $\pm0.3\%$ & $\pm0.2\%$ & --- \\
induced pseudoscalar, contact form or $g_3/2$ & $-1.0\%$ & $-0.3\%$ & --- \\
Monte Carlo, standard error & $0.9\%$ & $3.3\%$ & --- \\
\end{tabular}
\end{ruledtabular}
\end{table*}

\section{Generators}
\label{sec:generator}

The StrangeMC generator now carries the reference $f(E)$ of Table~\ref{tab:chain} in place
of the flat $15\%$. The flat value is $26$--$33\%$ low at $2$--$5\GeV$. It applies $f$ to its
own quasi-elastic hyperon cross section at $0.6$--$10\GeV$. Outside that range it keeps
the normalisation factor of the nearest end point rather than $f$. Below $0.6\GeV$ the two-body cross
section therefore follows the threshold of its own phase space, not that of the
quasi-elastic cross section. The denominator of $f$ is the strange quasi-elastic cross section per
nucleon without RPA quenching. The rate $f\sigma_{\rm QE}$ is shared among the seven
charge channels of \paperI: $\nubar\,pn\to\ell^+\Lam n$, $\ell^+\Sig^0n$ and
$\ell^+\Sig^-p$; $\nubar\,pp\to\ell^+\Lam p$, $\ell^+\Sig^0p$ and $\ell^+\Sig^+n$; and
$\nubar\,nn\to\ell^+\Sig^-n$. The $pp$ and $nn$ channels are thus shares of $f$, not
additions to it: the primary $\Sig^+$ is $1.1$--$1.2\%$ of the rate and the other
like-pair channels $3.6$--$3.7\%$. For the same reason the possible increase of $f$ from
the pair census (Sec.~\ref{sec:uncert}) is not applied.
On $pn$ pairs the two-body current gives $\sigma_\Lam/\sigma_{\Sig^0}=3.3$--$3.4$ at
$1$--$5\GeV$, $0.81$--$0.87$ times the quasi-elastic value. Its
$\sigma_{\Sig^-}/\sigma_{\Sig^0}$ is $1.2$ at $1\GeV$, against $2.0$ for quasi-elastic
production on a free neutron and proton. Two-body events therefore have fewer $\Sig^-$
than quasi-elastic events and a similar $\Lam$ fraction.
Summed over the pair types the composition is $\Lam\!:\!\Sig^0\!:\!\Sig^-=4.0\!:\!1\!:\!1.28$
at $0.6\GeV$ and $3.3$--$3.5\!:\!1\!:\!1.23$--$1.27$ at $1$--$5\GeV$. It inherits the
matching uncertainty: across the prescriptions $\sigma_\Lam/\sigma_{\Sig^0}$ is
$0.67$--$0.95$ of the quasi-elastic value at $1$--$5\GeV$ (\paperI). The generator
evaluates the composition at the pair-equivalent energy of each event, the beam energy at
which two nucleons at rest would have the invariant mass of the event
(Appendix~\ref{app:current}).

The generator samples are on argon, for SBND and DUNE in antineutrino and neutrino mode.
Each has $5\times10^4$ events that produce a strange hadron at the primary vertex, called
strange events below. They are generated with these channel weights, \textsc{Pythia}8
fragmentation~\cite{pythia83}, correlated nucleon pairs and the intranuclear cascade. They
use the carbon $f$ unchanged. The generator adds the quasi-elastic and
two-body rates, never their amplitudes, so the fractions below contain no interference.
In a simulated SBND antineutrino-mode (SBND-RHC) exposure~\cite{sbnd2025} the two-body
class is $10.8\%$ of all strange events, third after quasi-elastic production ($66.4\%$)
and associated production of a hyperon and a kaon ($16.7\%$). On the generator's approximation of the DUNE
antineutrino-mode flux~\cite{dune2020} (DUNE-RHC) it is $3.6\%$, because deep-inelastic strange
production supplies $60\%$ there. That value rests on the values of $f$ above $3\GeV$,
which are not predictions (Sec.~\ref{sec:results}). In the neutrino-mode samples the
wrong-sign antineutrinos give $1.7\%$ (SBND) and $0.2\%$ (DUNE).
The SBND-RHC two-body events have $\Lam/\Sig^0=3.76(14)$ and $\Sig^-/\Sig^0=1.22(6)$,
against $3.6$ and $1.22$ expected from the channel weights for bound nucleons at these
energies.

With the other channels fixed, a factor $r$ on the two-body rate moves its share $s$ to
$rs/[(1-s)+rs]$, so a $10\%$ change of $f$ moves the SBND-RHC share by $1.0$ percentage
point.
Table~\ref{tab:scenarios} applies this with $r=\langle f\rangle/0.169$ to the variations
of Sec.~\ref{sec:uncert}. The share runs from $8.8\%$ to $15.1\%$, and quasi-elastic
production stays the largest component ($63$--$68\%$). The two-body class would pass
associated production only for $r>1.54$, which no variation reaches.

\begin{table*}[t]
\caption{\label{tab:scenarios}Shares of the SBND-RHC strange sample under the variations
of Sec.~\ref{sec:uncert}, in the generator configuration of Sec.~\ref{sec:generator}.
Each row rescales the reference two-body rate by $r=\langle f\rangle/0.169$ (RHC
spectrum). The shares follow from $rs/[(1-s)+rs]$ with $s=10.8\%$ of all strange events.
The last column is the fraction of the kaon-less $\Sig^+$ tag made of primary $\Sig^+$
that leave the nucleus, $45r/(266+103r)$: $266$ quasi-elastic and $103$ two-body $\Sig^+$
in the tag, $45$ of them primary (Sec.~\ref{sec:experiments}).
All rows except the last use the carbon $f$. The type column follows the groups of
Table~\ref{tab:budget}.}
\tablefont
\begin{ruledtabular}
\raggedcells\begin{tabular}{p{5.6cm}lccc}
variation & type & $\langle f\rangle_{\rm RHC}$ & 2p2h share [\%] & primary $\Sig^+$/tag [\%] \\
\hline
FL reference, $\Delta=0.10\GeV^2$ & reference & $0.169$ & 10.8 & 12.2 \\
FL, $\Delta=0.046\GeV^2$ & definition & $0.217$ & 13.5 & 14.5 \\
FL, $\Delta=0.216\GeV^2$ & definition & $0.136$ & 8.9 & 10.4 \\
PV, $\epsilon=0.02\GeV^2$ & definition & $0.249$ & 15.1 & 15.9 \\
PV, $\epsilon=0.03\GeV^2$ & definition & $0.228$ & 14.0 & 15.0 \\
PV, $\epsilon=0.06\GeV^2$ & definition & $0.186$ & 11.8 & 13.1 \\
decuplet deleted, both sectors & diagnostic & $0.222$ & 13.7 & 14.7 \\
uniform correlation, $q_c=0.78\GeV$ & scenario & $0.170$ & 10.9 & 12.2 \\
RPA, $Q_S=Q_{NS}$ & convention & $0.202$ & 12.6 & 13.8 \\
extreme saturation & scenario & $0.164$ & 10.5 & 11.9 \\
$\kappa$ at $+20\%$ & input & $0.203$ & 12.7 & 13.9 \\
$\kappa$ at $-20\%$ & input & $0.135$ & 8.8 & 10.3 \\
C $\to$ Ar, $\times1.24$ (illustrative) & extrapolation & $0.210$ & 13.1 & 14.2 \\
\end{tabular}
\end{ruledtabular}
\end{table*}

\section{Signatures}
\label{sec:experiments}

Every two-body event carries a second nucleon from the pair. The two processes share
the hyperon final state, so an inclusive rate alone cannot separate them. They are statistically
separable through the recoil nucleon~\cite{kasturi2026}, within the limits
set by detector thresholds, final-state interactions and neutron invisibility. No
detector is simulated; the distributions are at generator level, in the argon samples of
Sec.~\ref{sec:generator}, which use the carbon $f$. The generator takes the channel rates
from \paperI\ and $f$ from this work. It draws the kinematics of each channel from a
simplified V$-$A matrix element on three-body phase space. The distributions of this section therefore depend on that choice and are not predictions of the current of \paperI.

\paragraph{Kaon-less $\Sigma^+$ tag.} Events with a $\Sig^+$ and no kaon are used to tag
hyperon charge exchange in the nucleus~\cite{companionPRD}, because one-body production
cannot make a $\Sig^+$. The $\Delta S=\Delta Q$ rule, by which the hadronic charge changes
with the strangeness, forbids it at any one-body weak vertex. The two-body
current evades the rule. On a $pp$ pair a charge exchange inside the pair produces a
primary $\Sig^+$ (\paperI), $1.1$--$1.2\%$ of the two-body rate
(Sec.~\ref{sec:generator}). In the SBND-RHC sample the two-body current supplies $28\%$ of
the kaon-less $\Sig^+$ tag. Primary $\Sig^+$, which owe nothing to charge exchange in the
nucleus, make up $12\%$ of it.
The tag holds $369$ $\Sig^+$ after final-state interactions,
$0.86\%$ of hyperon events. Of these, $266$ come from quasi-elastic and $103$ from
two-body events. Of the $103$, $45$ are primary $\Sig^+$ that leave the nucleus, $12.2\%$
($\pm1.8\%$) of the tag. A separate run of $3\times10^4$ two-body events gives a two-body
share of $30\%$ and a primary share of $11.3\%$; the DUNE-RHC flux gives $33\%$ and
$11.8\%$.
The other $58$ come from the cascade of two-body $\Lam$ and $\Sig$ hyperons. The tag rate
is therefore a mixture of charge exchange and two-body production. An analysis that
extracts the charge-exchange strength from it needs the two-body channel in its hyperon
source model.

\begin{figure*}[t]
\centering
\includegraphics[width=0.98\textwidth]{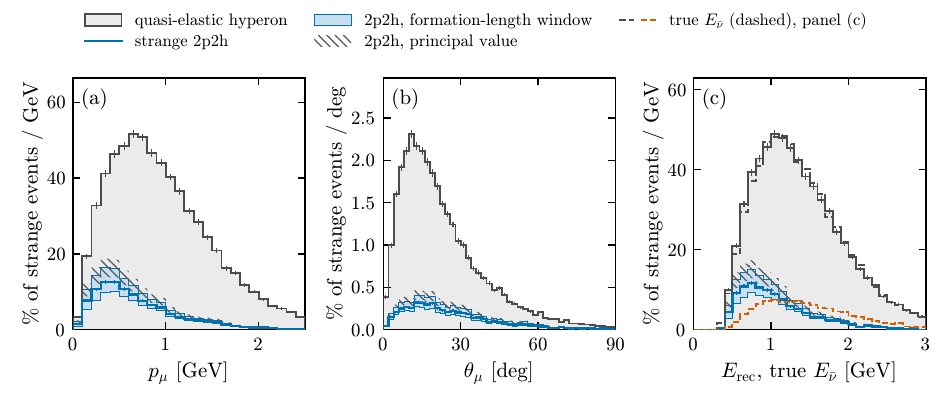}
\caption{\label{fig:dist}(a) Muon momentum, (b) muon angle and (c) reconstructed energy of
quasi-elastic hyperon events (filled histograms) and strange two-body events (lines) in
the SBND antineutrino-mode sample of Sec.~\ref{sec:generator} (argon with the carbon $f$,
$5\times10^4$ strange events). The distributions are in per cent of the strange sample per
unit of the variable, with statistical errors. $E_{\rm rec}$ is the
quasi-elastic reconstruction with $E_b=42\MeV$; dashed histograms in (c) are the true antineutrino energies, which for
the quasi-elastic events nearly coincide with $E_{\rm rec}$.
The bands rescale the two-body events by $f(E)$ in the formation-length window
$\Delta=0.046$--$0.216\GeV^2$ (band) and under the principal value,
$\epsilon=0.02$--$0.06\GeV^2$ (hatched).}
\end{figure*}

\paragraph{Energy reconstruction.} When the energy is reconstructed from the muon under
the quasi-elastic hypothesis, two-body events reconstruct low (Table~\ref{tab:bias}). In
the SBND-RHC sample they are $12\%$ low in the $0.6\GeV$ bin, the energy of the T2K and
Hyper-Kamiokande (Hyper-K) off-axis peak~\cite{t2k2011,hyperk2018}, and $22\%$ low in the $1\GeV$ bin. In the
$3\GeV$ bin of the DUNE-RHC sample they are $30\%$ low. Quasi-elastic
hyperon events are within $3\%$. As for non-strange
2p2h~\cite{martini2012,nieves2012,lalakulich2012}, the second nucleon takes energy that
the two-body formula does not account for. The bias grows with energy as more of the
energy transfer goes to it. The standard reconstruction for a target
nucleon at rest is
\[
  E_{\rm rec}=\frac{M_Y^2-(M_N-E_b)^2-m_\mu^2+2(M_N-E_b)E_\mu}
                   {2(M_N-E_b-E_\mu+p_\mu\cos\theta_\mu)},
\]
with $E_\mu$, $p_\mu$ and $\theta_\mu$ the muon energy, momentum and angle, $M_N$ the mean
nucleon mass, $E_b$ a binding energy and $M_Y$ the mass of the primary hyperon, taken as
identified.
The value $E_b=42\MeV$ is chosen so that the quasi-elastic events of this
simulation reconstruct without bias summed over the bins of Table~\ref{tab:bias}. Over the
whole SBND-RHC sample the two-body events then reconstruct $26\%$ low on average and the
quasi-elastic ones $0.6\%$ low (Fig.~\ref{fig:dist}(c)). The conventional argon value,
$30\MeV$, makes every bias up to $3$ points more negative.
A reconstruction that assumes a final neutron, as a standard antineutrino quasi-elastic
selection does,
lowers $E_{\rm rec}$ by a further $0.20\GeV$ or more for a $\Lam$ and $0.30\GeV$ for a
$\Sig$. In the three bins the quasi-elastic events then reconstruct $10$--$41\%$ low and
the two-body events $39$--$53\%$ low.

\begin{table*}[t]
\caption{\label{tab:bias}Fractional bias $E_{\rm rec}/E_{\rm true}-1$ of the
quasi-elastic energy reconstruction for quasi-elastic hyperon (one-particle--one-hole,
1p1h) and strange two-body
(2p2h) events, in three bins of true antineutrino energy and over the whole SBND-RHC
sample, with the mass of the primary hyperon and a target nucleon at rest. $E_b=42\MeV$
removes the 1p1h bias summed over the three bins; $30\MeV$ is the value conventional for
argon. $N$: number of events (DUNE-RHC on the generator's flux approximation); standard
error of the mean on the last digits in parentheses; 16--84\%: the central $68\%$ of the
events.}
\tablefont
\begin{ruledtabular}
\begin{tabular}{rllrcccc}
 & & & & \multicolumn{3}{c}{$E_b=42\MeV$} & $E_b=30\MeV$ \\
\cline{5-7}\cline{8-8}
bin [GeV] & flux & sample & $N$ & mean & median & 16--84\% & mean \\
\hline
0.55--0.65 & SBND-RHC & 1p1h & 1190  & $+1.7(3)\%$   & $+1.6\%$  & $-7.7$ to $+10.3\%$ & $-1.3(3)\%$   \\
           &          & 2p2h & 146   & $-11.8(8)\%$  & $-10.4\%$ & $-22.5$ to $-2.1\%$ & $-14.7(8)\%$  \\
0.93--1.07 & SBND-RHC & 1p1h & 3378  & $-0.7(2)\%$   & $-0.1\%$  & $-8.0$ to $+6.4\%$  & $-2.6(2)\%$   \\
           &          & 2p2h & 513   & $-21.6(6)\%$  & $-20.7\%$ & $-37.9$ to $-7.4\%$ & $-23.5(6)\%$  \\
2.7--3.3   & DUNE-RHC & 1p1h & 1440  & $-1.3(2)\%$   & $-0.5\%$  & $-4.2$ to $+2.1\%$  & $-2.0(2)\%$   \\
           &          & 2p2h & 330   & $-30.1(11)\%$ & $-28.9\%$ & $-52.1$ to $-8.2\%$ & $-30.7(11)\%$ \\
all        & SBND-RHC & 1p1h & 33188 & $-0.58(5)\%$  & $-0.2\%$  & $-6.8$ to $+5.6\%$  & $-2.21(5)\%$  \\
           &          & 2p2h & 5403  & $-25.8(2)\%$  & $-24.0\%$ & $-45.6$ to $-7.8\%$ & $-27.3(2)\%$  \\
\end{tabular}
\end{ruledtabular}
\end{table*}

\paragraph{Vertex protons.}
In a liquid-argon detector a proton from the primary vertex is tracked above a kinetic
energy of a few tens of MeV, while the proton of $\Lam\to p\pi^-$ starts at the displaced
decay vertex. Quasi-elastic production makes no nucleon at the weak vertex, so its
vertex protons come from the rescattering of the hyperon. The two-body current makes a
primary proton in $21\%$ of its events, mostly together with a $\Sig^-$
($\nubar\,pn\to\ell^+\Sig^-p$). In the SBND-RHC sample this proton has a median kinetic
energy of $0.21\GeV$. Its recoil neutrons knock out further protons. After final-state
interactions $51\%$ of the SBND-RHC two-body events have at least one proton above
$40\MeV$, against $19\%$ of the quasi-elastic hyperon events ($62\%$ and $21\%$ in the
DUNE-RHC sample). A $25\MeV$ threshold adds $3$--$5$ percentage points. Events with a
primary $\Lam$ or $\Sig^0$ and a hyperon that leaves the nucleus give the $V^0$ topology
of a displaced $\Lam\to p\pi^-$ decay. For these the fractions are $40\%$ and $16\%$
($52\%$ and $19\%$ in the DUNE-RHC sample). Requiring a vertex proton therefore enriches a $V^0$
sample in two-body events by a factor of $2.6$--$2.7$.

\paragraph{Neutron tagging.} In a water Cherenkov detector the recoil nucleon is
invisible if it is a neutron and usually below threshold if it is a proton, so a
two-body event looks like a quasi-elastic one. On $pn$ pairs the second nucleon is a
neutron whenever the hyperon is a $\Lam$ or a $\Sig^0$, which is $78\%$ of the two-body
rate (\paperI). Quasi-elastic production leaves no neutron at the weak vertex. Both
classes also make neutrons through $\Sig^-\to n\pi^-$, $\Lam\to n\pi^0$ and final-state
interactions, so the two-body channel appears as an excess of one neutron on top of these.
Neutron counting in Super-Kamiokande with gadolinium~\cite{beacom2004} and in
Hyper-K~\cite{hyperk2018} could separate it statistically, though not event by
event.
In the argon samples, after final-state interactions, $93\%$ of the two-body
events and $34\%$ of the quasi-elastic ones have at least one neutron.

\paragraph{Muon and hyperon kinematics.}
Two-body events have softer muons at larger angles (Fig.~\ref{fig:dist}):
$\langle p_\mu\rangle=0.71\GeV$ and $\langle\theta_\mu\rangle=26.8^\circ$, against
$1.00\GeV$ and $23.9^\circ$ for quasi-elastic hyperon events in the SBND-RHC sample.
The $\Lam$ that leave the nucleus, before detection thresholds, are harder in two-body
events: $\langle p_\Lam\rangle=0.50\GeV$ against $0.42\GeV$ in the SBND-RHC sample, which
spans the whole flux, and $0.75$ against $0.51\GeV$ in the DUNE-RHC sample. The hyperon-potential
analysis of Ref.~\cite{companionPRD} uses a single representative energy per beam. At that
single energy the SBND difference disappears ($-0.02\GeV$) and the DUNE one remains
($+0.25\GeV$). At SBND the difference therefore comes from averaging over the flux. The harder
spectrum changes the hyperon acceptance.

\paragraph{Threshold.} On a free $pn$ pair at rest the channel opens at
$E_{\nubar}=\delta+\delta^2/(2M_{NN})\simeq0.30\GeV$, with $\delta=M_\Lam+m_\mu-M_p$ and
$M_{NN}$ the mass of the pair. This is below the $0.33\GeV$ of quasi-elastic $\Lam$
production on a nucleon at rest, because the pair, twice as heavy as a nucleon, takes half
the recoil energy. It lies inside the GeV-scale accelerator fluxes
and the atmospheric antineutrino flux.

\paragraph{Size of the effect.} Quasi-elastic hyperon production is about $2.8\%$ of an
antineutrino charged-current sample at BNB energies on argon~\cite{companionPRD}. The
two-body current adds $\langle f\rangle=0.15$--$0.17$ of that, about $0.5\%$ of the
$\nubar$ charged-current sample ($0.6\%$ with the argon extrapolation). At that size, and
with a bias of $-12\%$ at $0.6\GeV$, the channel is not a leading systematic for T2K. The
generators that T2K and Hyper-K use do not contain it. The decays $\Lam\to p\pi^-$ ($64\%$) and
$\Sig^-\to n\pi^-$~\cite{pdg2024} add a charged pion, so some of these events enter
single-pion samples.

\paragraph{Nucleon-decay backgrounds.} The channel matters little for the atmospheric
antineutrino background to $p\to\bar\nu K^+$ in Super-Kamiokande~\cite{superk2014} and
JUNO~\cite{juno2023pdk}. It produces a hyperon and no kaon, and feeds that background
only through rare kaon production by the hyperon or its decay products in the nucleus.

\section{Conclusions}
\label{sec:summary}

\begin{enumerate}\setlength{\itemsep}{0pt}
\item \emph{Size.} On carbon, and under the transfer ansatz, the strange two-body cross section is $0.14$--$0.15$ of the
quasi-elastic hyperon cross section at $0.6$--$1\GeV$ and $0.20$--$0.22$ at $2$--$5\GeV$.
Averaged over the BNB antineutrino spectra it is $0.153$--$0.169$, close to the $15\%$
placeholder, which is $26$--$33\%$ low at $2$--$5\GeV$. StrangeMC now uses this
$f(E)$.

\item \emph{Consequences for experiments.} In the StrangeMC simulation, with the carbon
$f$ applied to argon, the two-body current supplies $28\%$ of the SBND kaon-less $\Sig^+$
tag; primary $\Sig^+$ that leave the nucleus are $12\%$ of it. The tag rate therefore does not
measure hyperon charge exchange alone. Two-body events reconstruct $12$--$30\%$ low under the
quasi-elastic hypothesis and carry a second nucleon: $51\%$ of them have a proton above
$40\MeV$ at the vertex, against $19\%$ of quasi-elastic events. They also contain fewer
$\Sig^-$ and have softer muons. Their $\Lam$ spectrum is harder in the flux-averaged
samples; at SBND the difference comes from the flux average and is absent at a fixed
energy. The class is about $11\%$ of the SBND antineutrino-mode
strange sample ($9$--$15\%$ across the variations) and $3.6\%$ of the DUNE one. These
distributions follow from the generator's effective two-body kinematics and are model
dependent.

\item \emph{Use.} A generator can apply $f(E)$ of Table~\ref{tab:chain}, with the
seven-channel composition, to its own quasi-elastic hyperon cross section. Generators can
vary $f$ over the range of the two matching prescriptions, $0.80$--$1.47$ times the reference, with
$\pm20\%$ on the non-strange input $\kappa$. Analyses of hyperon and kaon-less $\Sig^+$ samples
need the channel in their signal and background models.

\item \emph{Uncertainty and next step.} The largest evaluated uncertainty is the
separation of the two-body current from one-body production followed by rescattering:
across the explored prescriptions $f$ at $1\GeV$ ranges from $0.12$ to $0.22$. The non-strange input, the RPA
convention (up to $+26\%$) and the in-medium decuplet widths ($-35\%$ to $+15\%$) are of
similar size. Interference with one-body production is not calculated. An illustrative
coherence $|\rho|=0.1$ would change the rate at $1\GeV$ by $\pm0.08$ of the quasi-elastic
one. At $|\rho|\simeq0.2$ the change equals the two-body rate. Below $1\GeV$, which
carries $28\%$ of the RHC and $62\%$ of the wrong-sign Booster rate, $f$ rests on nuclear inputs held at
their $1\GeV$ values. It is too high there if $\kappa\,Q_{NS}$ keeps falling with
decreasing energy. The values above $3\GeV$ show the behaviour of the free-pair treatment
and are not predictions. The normalisation rests on the transfer ansatz
and on the Valencia calculation. The non-strange mirror reproduces the Valencia energy
dependence to $7$--$12\%$ at $2$--$10\GeV$. On argon $f$ is estimated at $9$--$36\%$
above carbon. Replacing the transfer ansatz and the range of matching prescriptions by a
calculation
requires the nuclear response of the current, including in-medium resonance absorption
($\Sst N\to YN$), three-nucleon terms and the interference with one-body production.
\end{enumerate}

\begin{acknowledgments}
Although the author is a member of the DUNE and SBND collaborations, all views
presented here are his and not those of the collaborations as a whole. This work was
supported by the Science and Technology Facilities Council (STFC) Lancaster EPP
Consolidated Grant 2025--2029: UKRI2846.
\end{acknowledgments}

\appendix

\section{Model details}
\label{app:current}

\paragraph{Couplings, dressings and regulators.} The current is built from the
lowest-order SU(3) chiral Lagrangian with $D=0.804$, $F=0.467$ and
$f_K=f_\pi=92.4\MeV$ (\paperI); the mirror is the substitution of
Table~\ref{tab:mirror}. The octet weak vertices carry dipoles
with $M_V=0.84\GeV$ and $M_A=1.0\GeV$~\cite{bernard2002}. The decuplet weak vertex carries
the vector form factors $C_3^V(0)=2.13$, $C_4^V(0)=-1.51$ and $C_5^V(0)=0.48$ with the
$Q^2$ dependence of Ref.~\cite{hnv2007} ($M_V=0.84\GeV$), and the axial form factor
$C_5^A(0)=1.2$ with a dipole of axial mass $0.95\GeV$. These values are for $n\to\Delta^+$
and are $\sqrt3$ times larger for $p\to\Delta^{++}$. The axial form factor $C_4^A$, which
the Adler model of Ref.~\cite{hnv2007} sets to $-C_5^A/4$, is omitted. The $\Sst$ form factors follow from
SU(3)~\cite{alam2012}. The decuplet strong couplings come from $\Gamma_{\Sst}=36\MeV$
and $\Gamma_\Delta=117\MeV$ with the isospin Clebsch--Gordan factor of each charge
channel; the decuplet propagators are off-shell Rarita--Schwinger propagators with these
free widths. Each exchanged meson carries a monopole with $\Lambda_\pi=1.3\GeV$.

\begin{table*}[t]
\caption{\label{tab:mirror}The $\DS=0$ mirror as a substitution in the $\DS=1$
current of \paperI. The Lagrangian, the regulators, the pair configuration, the matching
prescription and the momentum-transfer restriction are common to the two sectors. $f_1$,
$f_2$ and $g_1$ are the vector, weak-magnetism and axial couplings, and $\kappa_p$,
$\kappa_n$ the anomalous magnetic moments. Direct and crossed: $s$- and $u$-channel
resonance; on line 2: attached to the second nucleon.}
\tablefont
\begin{ruledtabular}
\raggedcells\begin{tabular}{p{3.0cm}p{6.2cm}p{6.6cm}}
ingredient & $\DS=1$ current & $\DS=0$ mirror \\
\hline
weak one-body vertex & $p\to\Lam$, $p\to\Sig^0$, $n\to\Sig^-$, SU(3) couplings $f_1$, $f_2$, $g_1$ of \paperI & $p\to n$: $f_1=-1$, $f_2=-(\kappa_p-\kappa_n)=-3.706$, $g_1=+(D+F)$ \\
decuplet poles & $\Sst$, $\Gamma=36\MeV$, direct and on line 2; crossed $\Delta(1232)$; form factors of Ref.~\cite{hnv2007} with SU(3) factors & $\Delta(1232)$, $\Gamma=117\MeV$, direct and crossed; form factors of Ref.~\cite{hnv2007} \\
final state & $\ell^+YN'$ & $\ell^+nn$ from a $pn$ pair, antisymmetrised, factor $1/2$ \\
\end{tabular}
\end{ruledtabular}
\end{table*}

\paragraph{Ratios.} $R$ and $\Xi$ are computed on the same targets, one $pn$ pair and
one proton and one neutron:
\begin{gather*}
  R=\frac{\sigma_{\nubar pn\to\ell^+\Lam n}+\sigma_{\nubar pn\to\ell^+\Sig^0n}+\sigma_{\nubar pn\to\ell^+\Sig^-p}}
         {\sigma_{\nubar pn\to\ell^+nn}},\\
  \Xi=\frac{\sigma_{\nubar p\to\ell^+n}}
           {\sigma_{\nubar p\to\ell^+\Lam}+\sigma_{\nubar p\to\ell^+\Sig^0}+\sigma_{\nubar n\to\ell^+\Sig^-}},
\end{gather*}
with the final neutrons of the mirror antisymmetrised with the identical-particle factor
$1/2$ and the one-body
cross sections on a free nucleon at rest. On carbon ($Z=N$) only the protons undergo
$\nubar p\to\ell^+n$, while every nucleon undergoes $\DS=1$ production. $\Xi$ is therefore
also
the ratio of the free-nucleon quasi-elastic cross sections of the nucleus; on argon it
is not (Appendix~\ref{app:argon}). For identical initial nucleons the amplitude is
$M(1,2)-M(2,1)$, where $M(1,2)$ is the amplitude with nucleon 1 on line 1. The rate per
pair of the census, which counts unordered pairs, is $\sum|M(1,2)-M(2,1)|^2$, summed over the
final and averaged over the initial spins.

\paragraph{Matching.} Both prescriptions of Sec.~\ref{sec:boundary} act on the
octet-baryon pole diagrams, which \paperI\ groups into weak-first, absorption-first and
kaon-exchange pole chains. They replace the denominator $x=p^2-M^2$ of every octet-baryon
propagator $(\slashed p+M)/x$ in these chains and leave the numerator unchanged:
\[
  \frac1x\;\to\;\frac{\theta(|x|-\Delta)}{x}\quad\text{(formation-length cut)},\qquad
  \frac1x\;\to\;\frac{x}{x^2+\epsilon^2}\quad\text{(principal value)}.
\]
Octet propagators carry no width; the decuplet propagators keep theirs and are not
regulated. The
substitution is made chain by chain in the amplitude, before the coherent sum is
squared, and identically in the two sectors. The principal-value regulator $\epsilon$ is a
matching scale that \paperI\ quotes at $0.03\GeV^2$. \paperI\ also gives the cost of each
prescription to the vector Ward identity. The generator applies the cut at the
current level and runs its cascade on an on-shell hyperon, so each pole chain is assigned
either to the two-body current or to the cascade, and nothing is counted twice.

\paragraph{Omitted terms and dressing.} The axial contact operator and its kaon-pole
partner are omitted together. Restoring the two at the Lagrangian coefficient lowers
$\sigma_\Lam/\sigma_{\Sig^0}$ by $6\%$ at $1\GeV$ and $2\%$ at $3\GeV$ and changes the
$\DS=1$ two-body rate by about $1\%$ (\paperI). Neither sector of $R$ has a three-nucleon
term. $\kappa$, however, contains the three-nucleon $C_{A3}$ piece of the in-medium
$\Delta$ self-energy~\cite{osetsalcedo1987}, shown separately as
three-particle--three-hole (3p3h) strength in Ref.~\cite{sobczyk2020}. Part of the Valencia shortfall of Sec.~\ref{sec:valencia} can
therefore be three-nucleon strength. The strange three-nucleon weak vertex has been
studied, with the $W$ exchanged between baryons, in the hypernuclear decay $\Lam NN\to NNN$~\cite{itonaga2023}.
Before form-factor dressing the vector current of \paperI\ satisfies its SU(3)-limit
Ward identity, and the dipole form factors of the reference calculation break it. A Ward-compatible dressing changes
the channel-summed $\DS=1$ pair rate by at most $2.8\%$ over $0.6$--$5\GeV$ and
$\sigma_\Lam/\sigma_{\Sig^0}$ by at most $4.3\%$ (\paperI). The mirror, and with it $R$,
has not been evaluated with that dressing.

\paragraph{Generator weights.} Each channel is weighted by its rate in \paperI\ over the
generator's own rate for that channel on two nucleons at rest. Both rates are taken at the
pair-equivalent energy of Sec.~\ref{sec:generator}, $E_{\rm
eq}=[E_N(E_N+E_{\nubar})-M^2]/M$ ($0.6\GeV$ when lower), with $M$ and $E_N$ the mass and
energy of the struck nucleon. A free nucleon at rest has
$E_{\rm eq}=E_{\nubar}$; for a bound nucleon $E_{\rm eq}$ lies $0.10$--$0.15\GeV$ below the
beam energy at $1\GeV$, which raises $\Lam/\Sig^0$ by $5$--$8\%$ at $0.7$--$1\GeV$.

\begin{figure*}[t]
\centering
\includegraphics[width=0.98\textwidth]{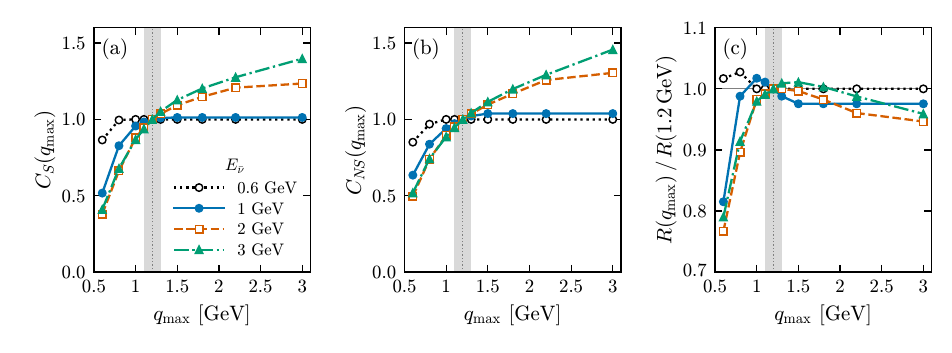}
\caption{\label{fig:qscan}Cumulative momentum-transfer scan at
$E_{\nubar}=0.6$, $1$, $2$ and $3\GeV$ ($2\times10^4$ phase-space points per energy).
(a) $C_S(q_{\max})$, the $\DS=1$ rate below $q_{\max}$ relative to the rate below the
adopted limit. (b) The same for the $\DS=0$ mirror.
(c) The ratio $R(q_{\max})/R(1.2\GeV)$. The dotted line marks the adopted $1.2\GeV$ limit
and the grey band its $1.1$--$1.3\GeV$ variation of Sec.~\ref{sec:uncert}.}
\end{figure*}

\paragraph{Momentum-transfer scan.} Figure~\ref{fig:qscan} shows the cumulative scan of
the restriction, $C_S(q_{\max})=\sigma_S(|\mathbf q|<q_{\max})/\sigma_S(|\mathbf
q|<1.2\GeV)$ and likewise $C_{NS}$ for the mirror. The restriction keeps the calculation
within the range of the Valencia calculation (Sec.~\ref{sec:mirror}) and is unrelated to
detector acceptance.

\paragraph{Interference among the classes.} Table~\ref{tab:interf} gives the coherent over
the incoherent sum of the diagram families (contact, meson in flight, octet poles, kaon
exchange and decuplet poles) in each sector. The deficit, one minus this ratio, grows for
$\DS=1$
from $14\%$ at $0.6\GeV$ to $21$--$22\%$ at $1$--$10\GeV$, and the $\DS=0$ deficit stays at
$8$--$10\%$. $R$ and every number derived from it are computed from the coherent sum.

\begin{table}[t]
\caption{\label{tab:interf}Ratio of the coherent to the incoherent sum of the
diagram-family cross sections in each sector. The last column is the quotient of the two,
the factor by which the interference changes $R$. Means of
three independent replicas of $2.5\times10^4$ phase-space points, with the Monte Carlo error on the
last digits in parentheses; the error of the quotient is from the replica-by-replica
quotient. The $10\GeV$ entries are the least converged.}
\tablefont
\begin{ruledtabular}
\begin{tabular}{rccc}
$E_{\nubar}$ [GeV] & $\DS=1$ & $\DS=0$ & $\DS=1/\DS=0$ \\
\hline
0.6  & $0.858(3)$  & $0.908(1)$  & $0.945(4)$  \\
1.0  & $0.786(2)$  & $0.897(1)$  & $0.876(3)$  \\
2.0  & $0.778(3)$  & $0.911(4)$  & $0.853(7)$  \\
3.0  & $0.779(4)$  & $0.917(3)$  & $0.850(1)$  \\
5.0  & $0.779(7)$  & $0.922(10)$ & $0.845(8)$  \\
10.0 & $0.790(2)$  & $0.911(25)$ & $0.869(23)$ \\
\end{tabular}
\end{ruledtabular}
\end{table}

\section{Inputs and variations}
\label{app:inputs}

\paragraph{Valencia inputs.} Table~\ref{tab:kappa} lists the antineutrino cross sections
of Ref.~\cite{gran2013} on carbon within $|\mathbf q|<1.2\GeV$ and the two ratios taken
from them. The quench $Q_{NS}$ is strongest at low energy, as expected of a
long-wavelength effect. Below $1\GeV$, where the table has no entry, $\kappa$ and $Q_{NS}$
are held at their $1\GeV$ values; at $1.5\GeV$ they are interpolated linearly between the
$1$ and $2\GeV$ entries.

\begin{table}[t]
\caption{\label{tab:kappa}Antineutrino 2p2h and quasi-elastic cross sections on carbon
from Table~I of Ref.~\cite{gran2013} (local Fermi gas, $|\mathbf q|<1.2\GeV$), in the
units of that table; only their ratios enter. The last column is the shape against which
the mirror is compared in Fig.~\ref{fig:valencia}.}
\tablefont
\begin{ruledtabular}
\begin{tabular}{rcccccc}
$E_{\nubar}$ [GeV] & $\sigma_{2p2h}$ & $\sigma_{\rm QE}^{\rm LFG+RPA}$ & $\sigma_{\rm QE}^{\rm LFG}$ & $\kappa$ & $Q_{NS}$ & $\sigma_{2p2h}(E)/\sigma_{2p2h}(1\GeV)$ \\
\hline
1  & 0.459 & 1.56 & 1.96 & 0.294 & 0.80 & 1 \\
2  & 0.887 & 2.68 & 3.03 & 0.331 & 0.88 & 1.932 \\
3  & 1.07  & 3.26 & 3.55 & 0.328 & 0.92 & 2.331 \\
5  & 1.24  & 3.83 & 4.05 & 0.324 & 0.95 & 2.702 \\
10 & 1.38  & 4.31 & 4.47 & 0.320 & 0.96 & 3.007 \\
\end{tabular}
\end{ruledtabular}
\end{table}

\begin{table}[t]
\caption{\label{tab:rpascen}Illustrative strange-sector quenching options: $f$ under
three treatments of RPA quenching in the strange quasi-elastic denominator, at the
reference matching, from the replica means of Table~\ref{tab:chain}.
The Monte Carlo error is $0.9\%$ at $1$ and $2\GeV$ and $3\%$ at $3$--$10\GeV$. The
last two columns are the Booster averages (FHC\,ws: wrong-sign $\bar\nu_\mu$ component).}
\tablefont
\begin{ruledtabular}
\begin{tabular}{lccccccc}
$Q_S$ & $1\GeV$ & $2\GeV$ & $3\GeV$ & $5\GeV$ & $10\GeV$ & $\langle f\rangle_{\rm RHC}$ & $\langle f\rangle_{\rm FHC\,ws}$ \\
\hline
$1$ (adopted) & 0.150 & 0.202 & 0.211 & 0.223 & 0.213 & 0.169 & 0.153 \\
$\sqrt{Q_{NS}}$ & 0.168 & 0.214 & 0.220 & 0.229 & 0.217 & 0.185 & 0.169 \\
$Q_{NS}$ (equal quenching) & 0.189 & 0.228 & 0.229 & 0.235 & 0.221 & 0.202 & 0.188 \\
\end{tabular}
\end{ruledtabular}
\end{table}

\paragraph{Variations.} Tables~\ref{tab:boundary} and~\ref{tab:src} give the matching and
correlation variations of Sec.~\ref{sec:uncert}. The Oset--Toki--Weise function
$g(r)=1-j_0(q_cr)$, $q_c=0.78\GeV$~\cite{oset1982}, subtracts from each exchange the same
exchange at $|\mathbf k|^2\to|\mathbf k|^2+q_c^2$, with $\mathbf k$ the exchanged-meson
momentum, and adds the induced spin--spin contact
(\paperI). At $q_c=0.60$ and $1.00\GeV$ each current falls to $0.20$--$0.24$ and
$0.42$--$0.46$ of its unscreened value, respectively, and $R$ stays within $1.4\%$. The Landau--Migdal
contact $g'=0.7$~\cite{oset1982,nieves2011} is added as $g'/|\mathbf k|^2$ to the
pion propagator $F_\pi(k^2)/(k^2-m_\pi^2)$, with $F_\pi$ the monopole, with no form factor on the
contact, on the exchanged pions only. Apart from the $\eta$, the mirror exchanges only
pions. The $\DS=1$ current also exchanges kaons, which the contact leaves unscreened. The
contact therefore suppresses the mirror more and raises $R$ by up to a factor $3.0$ at
$1$--$5\GeV$. At $3\GeV$ it gives
$R=0.131(3)$, $2.5$ times $\tan^2\theta_C$, against the unscreened $0.043$. The factor is
an artefact of screening the two sectors differently and is not propagated. Deleting the
decuplet class from both coherent sums is a diagnostic of its weight (two seeds of $2\times10^4$ phase-space points, common random numbers). It lowers the $\DS=1$ strength by $14\%$ at
$0.6\GeV$ and $35$--$55\%$ at $1$--$5\GeV$, and the $\DS=0$ strength by $42\%$ at
$0.6\GeV$ and $51$--$63\%$ at $1$--$5\GeV$. Because the class weighs more in the mirror, $R$ rises by a factor $1.49$ at
$0.6\GeV$, where the $\Sst$ is close to threshold, and $1.19$--$1.32$ at $1$--$5\GeV$.

\begin{table}[t]
\caption{\label{tab:boundary}$f$ under the two matching prescriptions. The last column,
$f/f_{\rm ref}$, equals $R$ in the given prescription over $R$ at the reference cut, paired by seed and
averaged over two seeds of $2\times10^4$ phase-space points. Each $f$ entry is that ratio times
the reference $f$ of Table~\ref{tab:chain}. Half the difference between the two seeds is at most $3.4\%$ at
$1\GeV$ and at most $10\%$ at $3$ and $5\GeV$. The cut is varied over the
formation-length window of \paperI; the principal value carries no cut, and its
$\epsilon=0.02\GeV^2$ row lies below the regulator at which \paperI\ quotes the
principal value. All rows are alternative definitions of the current--cascade
separation; the three cut rows are different scales of one definition.}
\tablefont
\begin{ruledtabular}
\begin{tabular}{lcccc}
prescription ($\GeV^2$) & $E=1\GeV$ & $E=3\GeV$ & $E=5\GeV$ & $f/f_{\rm ref}$ ($1$/$3$/$5\GeV$) \\
\hline
cut, $\Delta=0.046$ & 0.196 & 0.264 & 0.267 & 1.30 / 1.26 / 1.20 \\
cut, $\Delta=0.10$ (reference) & 0.150 & 0.211 & 0.223 & 1 \\
cut, $\Delta=0.216$ & 0.119 & 0.180 & 0.203 & 0.79 / 0.86 / 0.91 \\
principal value, $\epsilon=0.02$ & 0.222 & 0.297 & 0.294 & 1.48 / 1.41 / 1.32 \\
principal value, $\epsilon=0.03$ & 0.203 & 0.270 & 0.271 & 1.35 / 1.28 / 1.22 \\
principal value, $\epsilon=0.06$ & 0.166 & 0.226 & 0.230 & 1.10 / 1.07 / 1.03 \\
\end{tabular}
\end{ruledtabular}
\end{table}

\begin{table}[t]
\caption{\label{tab:src}$R/R_{\rm ref}$ under the two correlation treatments, at the
reference cut, from four replicas of $2\times10^4$ phase-space points run with the seeds of the
reference replicas of Table~\ref{tab:chain}. The Monte Carlo error on the last digits,
in parentheses, is the standard error of the replica-by-replica quotient. Uniform: the
correlation function on every exchanged meson, $q_c=0.78\GeV$, a physics scenario.
Pion only: the Landau--Migdal contact, $g'=0.7$, which screens some exchanges and not
others; a diagnostic, not propagated.}
\tablefont
\begin{ruledtabular}
\begin{tabular}{rcc}
$E_{\nubar}$ [GeV] & uniform & $\pi$ only \\
\hline
0.6  & 1.008(1)  & 1.25(7)  \\
0.8  & 1.010(3)  & 1.63(7)  \\
1.0  & 1.002(2)  & 1.86(8)  \\
1.5  & 1.002(1)  & 2.40(15) \\
2.0  & 1.009(4)  & 2.73(14) \\
3.0  & 1.010(2)  & 3.04(16) \\
5.0  & 1.008(6)  & 3.03(19) \\
10.0 & 0.978(24) & 3.62(28) \\
\end{tabular}
\end{ruledtabular}
\end{table}

\subsection{Carbon to argon}
\label{app:argon}

With the transfer ansatz on each nucleus the non-strange quasi-elastic cross section
cancels between $\kappa\,Q_{NS}$ and $\Xi$ [Eq.~\eqref{eq:chain}], and per nucleon
\[
  \frac{f_{\rm Ar}}{f_{\rm C}}
  =c_{\rm Ar/C}\,
   \frac{[\sigma^{\DS=0}_{2p2h}/A]_{\rm Ar}}{[\sigma^{\DS=0}_{2p2h}/A]_{\rm C}}\,
   \frac{[\sigma^{\DS=1}_{\rm QE}/A]_{\rm C}}{[\sigma^{\DS=1}_{\rm QE}/A]_{\rm Ar}} . \]
The three factors are the change $c_{\rm Ar/C}$ of the strange channel sum per $pn$ pair
between the two pair censuses, the non-strange 2p2h scaling, and the change of the
one-body denominator. Final-state
interactions act through the cascade, not on $f$, which is defined at the primary vertex. The
censuses have $pp\!:\!pn=1\!:\!18$ on both nuclei, and $nn\!:\!pp=1.51$ on argon against
$1$ on carbon. Through $nn\to\Sig^-n$ this gives
$c_{\rm Ar/C}=1.004$--$1.006$ at $0.6$--$5\GeV$, which is not applied. The strange
quasi-elastic denominator, weighted by the proton and neutron numbers ($18\!:\!22$ against
$6\!:\!6$) with the one-body cross sections of \paperI, scales by $0.957$ per nucleon. The numerator follows two
published routes (Table~\ref{tab:argon}). In the model that $\kappa$ is taken from, the
2p2h cross section grows faster than $A$ once $\Delta$ absorption is included: per nucleon
it is $1.2$ times larger on calcium than on carbon~\cite{gran2013}. Calcium to argon is a
pair-counting factor $0.99$~\cite{schwehr2016}, giving $1.2\times0.99/0.957\simeq1.24$. The microscopic
$^{40}$Ar calculation of Ref.~\cite{martinez2025} gives per-nucleon transverse 2p2h
responses of $1.04$--$1.30$ times those of $^{12}$C over $|\mathbf q|=0.5$--$2\GeV$. After
the same $0.957$ this gives $f_{\rm Ar}/f_{\rm C}\simeq1.09$--$1.36$. A phase-space scaling law
reproduces such responses across the periodic table to about
$10\%$~\cite{Martinez-Consentino:2026yxv}. The SBND fractions of Sec.~\ref{sec:generator}
use the carbon $f$, and Table~\ref{tab:scenarios} carries the $\times1.24$ row.

\begin{table*}[t]
\caption{\label{tab:argon}Construction of the carbon-to-argon transfer factor of $f$, an
illustrative extrapolation; all other values of $f$ are for carbon. Neither route
calculates the antineutrino 2p2h cross section on argon within $|\mathbf q|<1.2\GeV$. Route 1 is for
neutrinos and the full response, route 2 for a response function at fixed
$|\mathbf q|$. The strange quasi-elastic scaling is computed from the one-body cross
sections of \paperI; the non-strange quasi-elastic rate cancels in $f$ and does not
enter. RMF: relativistic mean field; RFG: relativistic Fermi gas; $R^T$: transverse
response.}
\tablefont
\begin{ruledtabular}
\raggedcells\begin{tabular}{p{1.8cm}p{2.8cm}p{2.3cm}p{1.9cm}p{2.9cm}p{1.9cm}c}
route & source & quantity scaled & per nucleus / per nucleon & $\DS=1$ QE scaling & range & $f_{\rm Ar}/f_{\rm C}$ \\
\hline
1, Valencia $A$-dependence & Ref.~\cite{gran2013} Sec.~III\,D: LFG with in-medium $\Delta$, neutrino & $\sigma_{2p2h}$, Ca/C $=4.0$ & per nucleus ($1.2$ per nucleon); Ca$\to$Ar $\times0.99$ ($pn$)~\cite{schwehr2016} & per nucleon, $18\!:\!22$ vs $6\!:\!6$: $0.957$ & $|\mathbf q|<1.2\GeV$, $1$--$10\GeV$ & $1.24$ \\
2, low & Ref.~\cite{martinez2025}, RMF, transverse 2p2h response at the $\Delta$ peak & $R^T$, Ar/C $=1.04$ & per nucleon & $0.957$ & $|\mathbf q|=0.5\GeV$ & $1.09$ \\
2, high & same & $R^T$, Ar/C $=1.30$ & per nucleon & $0.957$ & $|\mathbf q|=2\GeV$ & $1.36$ \\
not used & Ref.~\cite{martinez2025}, RFG & $R^T$, Ar/C $=1.33$--$1.53$ & per nucleon & $0.957$ & $0.5$--$2\GeV$ & $1.39$--$1.60$ \\
\end{tabular}
\end{ruledtabular}
\end{table*}

\section{Reproducibility}
\label{app:repro}

Every number comes from the program of \paperI\ and the StrangeMC generator, available from the author on request
with a manifest that maps each table entry and figure to its command, seed and output
file. The generator samples of Sec.~\ref{sec:generator} use seed $2025$.
Table~\ref{tab:appbench} gives the values at $1\GeV$ against which a rebuild is checked.

\begin{table}[t]
\caption{\label{tab:appbench}Reproduction check at $E_{\nubar}=1\GeV$ and the reference
treatment, from unrounded inputs. The symmetry residuals and one-body benchmarks of the
current are in \paperI.}
\tablefont
\begin{ruledtabular}
\begin{tabular}{lc}
quantity & value \\
\hline
$R$, four replicas & $0.03593(33)$ \\
$\Xi$ (unquenched), four replicas & $17.865(8)$ \\
$\kappa=0.459/1.56$, Ref.~\cite{gran2013} & $0.2942$ \\
$Q_{NS}=1.56/1.96$, Ref.~\cite{gran2013} & $0.7959$ \\
$f=R\,\kappa\,\Xi\,Q_{NS}$ & $0.1503$ \\
\end{tabular}
\end{ruledtabular}
\end{table}

\bibliography{refs}

\end{document}